\documentclass{article}

\usepackage[preprint]{neurips_2024}

\usepackage[T1]{fontenc}    
\usepackage{hyperref}       
\usepackage{url}            
\usepackage{booktabs}       
\usepackage{amsfonts}       
\usepackage{nicefrac}       
\usepackage{microtype}      
\usepackage{xcolor}         

\usepackage{times}

\usepackage{soul}
\usepackage{url}
\usepackage[small]{caption}
\usepackage{graphicx}
\usepackage{amsmath}
\usepackage{amssymb}
\usepackage{bm}
\usepackage{booktabs}
\usepackage{threeparttable}
\usepackage{tabularx}
\usepackage{multirow,multicol, array}
\usepackage[linesnumbered,ruled,vlined]{algorithm2e}
\usepackage{color}
\usepackage{booktabs} 
\usepackage{threeparttable} 
\usepackage{siunitx} 
\usepackage{wrapfig}

\newcommand{\OSDEC}{\textit{OSDEC}}

\title{An Advanced Reinforcement Learning Framework for Online Scheduling of Deferrable Workloads in Cloud Computing}

\author{%
  Hang Dong\thanks{Equal contribution.} \\
  Microsoft \\
  \And
  Liwen Zhu\normalfont\textsuperscript{*}\thanks{Work done while internship at Microsoft Research Asia (MSRA), Beijing, China.} \\
  Peking University \\
  \And
  Zhao Shan \\
  Tsinghua University \\
  \And
  Bo Qiao \\
  Microsoft \\
  \And
  Fangkai Yang \\
  Microsoft \\
  \And
  Si Qin \\
  Microsoft \\
  \And
  Chuan Luo \\
  Beihang University \\
  \And
  Qingwei Lin \\
  Microsoft \\
  \And
  Yuwen Yang \\
  Microsoft \\
  \And
  Gurpreet Virdi \\
  Microsoft \\
  \And
  Saravan Rajmohan \\
  Microsoft \\
  \And
  Dongmei Zhang \\
  Microsoft \\
  \And
  Thomas Moscibroda \\
  Microsoft \\
}

\begin{document}

\maketitle

\begin{abstract}
Efficient resource utilization and perfect user experience usually conflict with each other in cloud computing platforms. Great efforts have been invested in increasing resource utilization but trying not to affect users' experience for cloud computing platforms. In order to better utilize the remaining pieces of computing resources spread over the whole platform, deferrable jobs are provided with a discounted price to users. For this type of deferrable jobs, users are allowed to submit jobs that will run for a specific uninterrupted duration in a flexible range of time in the future with a great discount. With these deferrable jobs to be scheduled under the remaining capacity after deploying those on-demand jobs, it remains a challenge to achieve high resource utilization and meanwhile shorten the waiting time for users as much as possible in an online manner. In this paper, we propose an online deferrable job scheduling method called \textit{Online Scheduling for DEferrable jobs in Cloud} (\OSDEC{}), where a deep reinforcement learning model is adopted to learn the scheduling policy, and several auxiliary tasks are utilized to provide better state representations and improve the 
performance of the model. With the integrated reinforcement learning framework, the proposed method can well plan the deployment schedule and achieve a short waiting time for users while maintaining a high resource utilization for the platform.
The proposed method is validated on a public dataset and shows superior performance.  
\end{abstract}

\section{Introduction}
Cloud computing has emerged as a powerful paradigm that enables access to a shared pool of computing resources through the Internet \cite{li2016negotiation}. The infrastructures, platforms, and resources (e.g., CPU, memory, storage, etc.) are provided in the form of virtual machines (VM) that are configurable for different users by the cloud service providers such as Amazon Web Service, Microsoft Azure, Google Cloud and so on. 
To support customers' growing demand for cloud computing services and better utilize the resources from the platform, it remains a great challenge to achieve high utilization of the computing resources with a strictly high service level at the same time \cite{mesbahi2018reliability}.


In order to better utilize the valley hours of the on-demand cloud computing workload, cloud platforms offer a type of deferrable VM products that can pre-collect the job requests with the valid time window in the near future apart from the basic job configuration information. 
However, it is still not trivial to intelligently schedule the deferrable jobs due to the following challenges. 
First, the scheduling task needs to be done online, with new job requests coming and the scheduling plan updating. Even the optimal static scheduling plan can turn to non-optimal after updating the job pool and the available capacity.  Moreover, it is quite hard to utilize pre-collected jobs well to achieve a better plan than the real-time job schedules. Last but not least, other issues such as users' experience should also be considered while improving the resource utilization, making the problem a multi-objective optimization problem in nature and hard to solve.

There have been a variety of heuristic methods proposed for different types of job scheduling problems, such as fair scheduling \cite{ghodsi2011dominant}, first-in-first-out \cite{Narwal2018FIFO}, and simple packing strategies \cite{grandl2014multi}. However, due to the over-simplified rule behind these heuristic methods, they cannot work consistently well in a dynamic complex system such as a cloud computing platform. Moreover, the heuristic methods do not consider the information of the pre-collected jobs in the scheduling process. The problem of job scheduling has also been studied as an optimization problem in the offline setting \cite{dongpredictive21}, but solving such optimization problem is time-consuming and cannot preserve the optimality in the online setting. 
With the fast advance in reinforcement learning (RL) models, researchers have found that the online decision setting of RL is naturally suitable for such an online scheduling scenario \cite{mao2016resource}. 
Nevertheless, current RL methods for online job scheduling in cloud computing are dealing with the real-time scheduling task rather than learning a policy that can well utilize the information of pre-collected jobs \cite{zhou2021deep}.

In this paper, we propose a solution called \textit{Online Scheduling for DEferrable jobs in Cloud} (\OSDEC{}), which is a reinforcement learning framework  for learning the online scheduling policies with pre-collected workloads and capacity information over the cloud computing platform.  With the learned scheduling model, the platform can properly balance the improvement of the overall utilization of the platform and the waiting time for the deferrable jobs. 


    
In summary, we make the following key contributions in this paper:

1) For the problem of online scheduling of deferrable jobs, we provide an effective formulation for this setting considering both the resource utilization and the average waiting time in the goal;

2) We propose \OSDEC{}, a novel reinforcement learning framework designed for the deferrable job online scheduling problem, which can effectively incorporate the information of pre-collected jobs to learn the scheduling policy;

3) Inspired by the benefits brought by auxiliary tasks on representation learning, we propose a series of auxiliary tasks that can help with extracting useful features required to make the scheduling plan and incorporate these tasks in the reinforcement learning framework;


4) The performance of \OSDEC{} is validated with extensive experiments for the deferrable job scheduling problem in real public industrial datasets by comparing it with state-of-the-art scheduling methods. 

\section{Related works}
The static job scheduling problem is a classic discrete optimization problem that has been proved to be NP-hard \cite{Fanjul2017}. Solving such static job scheduling problems often resorts to heuristics \cite{Fleszar2018,dongpredictive21}. Heuristic methods are still applicable for the online scheduling scenario where new jobs keep coming before fulfilling the original scheduling plan.
The commonly used heuristics include first-in-first-out (FIFO) \cite{Narwal2018FIFO}, shortest-job-first (SJF), highest-response-ratio-next (HRRN) \cite{latip2011highest} and fairness \cite{isard2009quincy}.  Three heuristics are combined in  \cite{grandl2014multi}, including best packing, shortest remaining job time, and fairness, to improve both the cluster efficiency and the average job completion time while simultaneously achieving good fairness. However, due to the simple static rule behind these heuristic methods, they cannot work consistently well in a dynamic complex system in cloud computing. Moreover,  these heuristic methods ignore the pre-collected job information for future deployment while doing the scheduling task.


There are also some early attempts to adopt reinforcement learning (RL) methods into online job scheduling problems due to the natural online decision-making setting of RL \cite{strens2006combining}. One of the most representative works among them is DeepRM \cite{mao2016resource}, which adopts the REINFORCE algorithm \cite{williams1992simple} into the jobs scheduling problem, representing the state space with distinct images and training the policy with convolutional neural networks (CNN).
 To improve the training efficiency, \cite{guo2020cloud} combines imitation learning into the DeepRM model to accelerate the policy search. However, these works have not considered the deferrable jobs to be deployed in the future while doing the scheduling. 
 
 


Auxiliary tasks that have different but relevant goals from the major task have been shown to be effective both in improving learning efficiency and converging to better policies for reinforcement learning models because they can achieve better representation learning results \cite{Lin2019NIPS,lyle_effect2021,kumar2021implicit}. Various auxiliary tasks are adopted to boost the performance of RL in gaming tasks in Atari games \cite{kartal2019terminal} and visual robotic manipulation tasks simulated in MuJoCo \cite{Lin2019NIPS}. However, not many works have further extended auxiliary tasks to other scenarios suitable for reinforcement learning, including the job scheduling scenario.

\begin{figure}[t]
    \centering
    \includegraphics[width=0.85\textwidth,trim={0pt 0pt 0pt 0pt},clip]{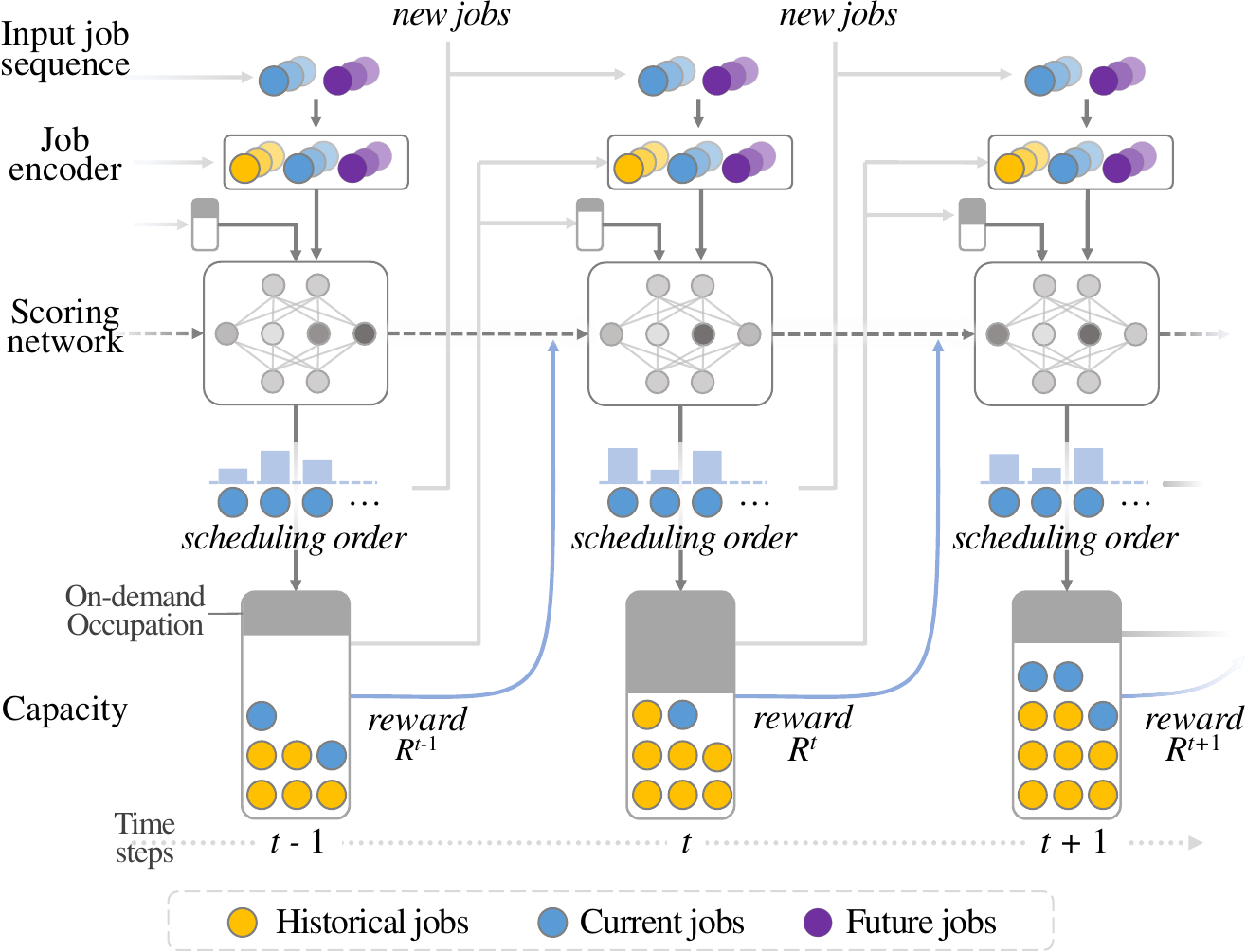}
    \caption{Overall design of \OSDEC{}.}
    \label{fig:algorthm}
\end{figure}

\section{Problem Formulation}
We consider the problem of deferrable job scheduling similar to the setting in \cite{dongpredictive21}, but have modified it into an online manner. Instead of a fixed set of pre-collected job requests to be scheduled, we consider the dynamic collection of job requests that allows new job requests to be submitted to the system while running. 
\subsection{Definitions on job requests and capacity}
The dynamic collection of job requests up to time $T$ are denoted as $B(T) = \{ b_1, \ldots, b_N \}$, each job request including basic requirements for the job to run. Specifically, for a job request $b_i \in B(T), i \in \{1, \ldots, N\}$, it can be represented as $b_i = (c_i, d_i, e_i, l_i, g_i)$, where $c_i$ is the requested resource capacity for job $i$,  $d_i$ is the duration for the job, $e_i$ and $l_i$ are the earliest start time and latest start time for the job respectively, and $g_i \leq e_i$ for $i = 1, ..., N$ is the submission time for the job, where we assume each job should be submitted before its earliest start time without loss of generality. In practice, the most valuable computing resource is the CPU cores in cloud computing \cite{dongpredictive21}, so in this paper, we use CPU cores to represent the requested resource for the job requests. Including any other type of resources up to the specific application is also easily adaptable from this setting.

The decision variables are denoted as $\{X_{it}\}$, $i = 1,...,N$, $t = 1, ..., T$, where $X_{it} = 1$ if job request $i$ is scheduled to start at time $t$, and $X_{it} = 0$ otherwise. We also denote the deployment time of each job $i$ as $t_i$, and by the definition $e_i \leq t_i \leq l_i$. Note that in this dynamic setting, only the ``submitted'' jobs are exposed to the system at each time step $t$: $B(t) = \{b_i| g_i \leq t\} \subseteq B(T)$, and this set of jobs keeps changing at each time step.

From the platform perspective, the value for each computing job is generally proportional to its occupied resource and its running time. Therefore, we use the CPU cores multiplied by the duration to represent the revenue brought by fulfilling each job request: $r_i = c_i d_i$ for $i \in \{1,...,N\}$. Moreover, even though each of these deferrable job requests has a deployment time window between the earliest start time and the latest start time, users would prefer the platform that can deploy their jobs as soon as they become valid in the real application scenario. Therefore, we also calculate the time length from the earliest start time till the actual deployment time as the delay for each job: 
$p_i = t_i - e_i$ for $i\in \{1,...,N\}$. 

The available capacity for those deferrable jobs is the remaining capacity obtained by subtracting the capacity occupied by on-demand jobs from the total capacity of the whole platform. Due to the uncertainty on future on-demand jobs, the future available capacity is also uncertain, and thus any scheduling plan would have the risk of exceeding the available capacity as long as there are deployed jobs. This violation of the available capacity limit at time step $t$ is 
$v_t = \max\left(0, \sum_{i=1}^N \sum_{t'=t-d_i}^t c_i X_{it'} - C_t\right)$
where $C_t$ is the available remaining capacity at time $t$.

\subsection{Problem formulation}
The online deferrable job scheduling problem considered in this paper is to decide which of the jobs should be deployed at each time $t$ so as the total occupied capacity should try not to exceed the available capacity at any time \cite{dongpredictive21} and try to make the total time delay for the deployed jobs as short as possible \cite{liu2017hierarchical}.  Therefore, we can integrate these actual requirements on the scheduling plan and represent the problem in the following form:

\begin{equation}\label{eq.our_problem}
\begin{aligned}
\textrm{maximize} & \sum_{i=1}^{N}\sum_{t=1}^TX_{it} \cdot ( r_i - \omega_1  p_i)  - \omega_2\sum_{t=1}^Tv_t \\
\text{s.t.}  &\sum_{t=1}^T X_{it}\le 1, \text{for } 1\le i\le N\\
            &\sum_{t< e_i \vee t>l_t}X_{it} = 0, \text{for } 1\le i\le N\\
          &X_{it}\in \{0,1\}, \text{for } 1\le i\le N, 1\le t\le T\\
\end{aligned}
\end{equation}
where $\omega_1$ and $\omega_2$ are the coefficient parameters that penalize the delay of jobs and the violation of available capacity respectively. From this offline optimization problem formulation in (\ref{eq.our_problem}), we can further modify it in the online setting by making the decisions $\{X_{it}\}$ sequentially for $t = 1,..., T$ and allowing the job set to add new jobs as well. For this dynamic setting, we use a reinforcement learning model to schedule the jobs as a sequential decision-making task and maximize the expected reward as an online surrogate of the goal in (\ref{eq.our_problem}), which will be illustrated in detail in the next section.

\section{Proposed Model}\label{sec.ppo_sorting}
In this section, we introduce the proposed \OSDEC{} model, which includes the reinforcement learning model design, the auxiliary tasks module, and the training process of the model. The overall design of the proposed method is shown in Figure \ref{fig:algorthm}. For each time step, there are several job requests submitted to the system and join the job pool. The available capacity, which is obtained by subtracting the capacity occupied by on-demand jobs from the total capacity, can be collected for the system. The proposed model will score the jobs and output the jobs that are chosen to be deployed at the current time. The reward for the scheduling plan is then calculated from the system as an indicator of how well the plan is, and the policy can be updated periodically with the latest real trajectory.  The details of this model will be explained in the following content.

\subsection{Reinforcement Learning Based Scheduling}

\begin{figure}[t]
    \centering
    \includegraphics[width=0.7\textwidth]{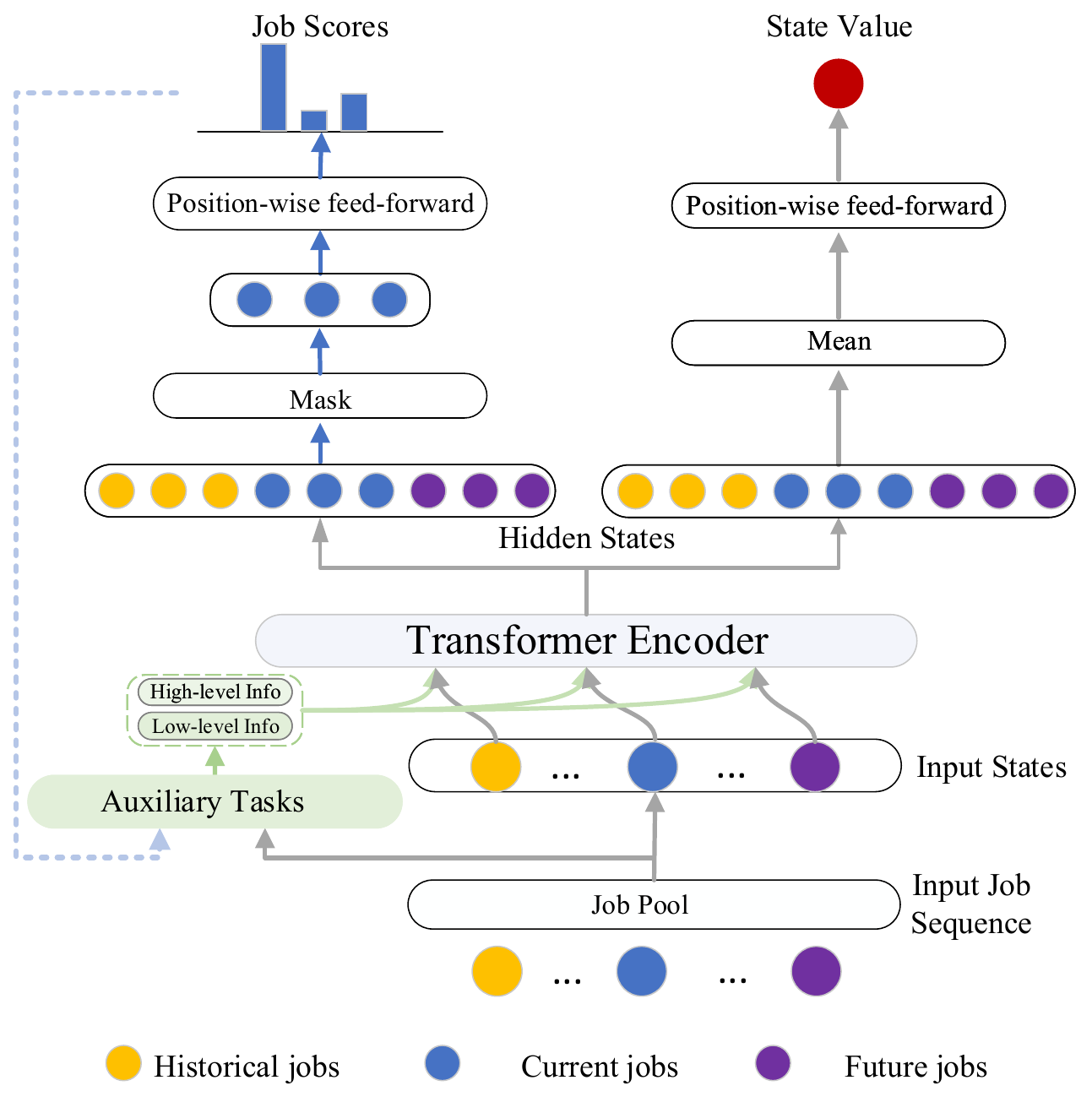}
    \caption{Overall network architecture of the proposed \OSDEC{} method.}
    \label{fig:policy_net}
\end{figure}

\subsubsection{State}
The explicit information that is available at the $t^{\text{th}}$ time step includes the submitted jobs $B(t)$ and the remaining available capacity $C_t$ during this time step. 
To better incorporate the available information into the states in the RL model, we separate the jobs in $B(t)$ in three distinct subsets: $B^{\text{his}}(t)$, $B^{\text{cur}}(t)$ and $B^{\text{fut}}(t)$. Among them, $B^{\text{his}}(t)$ represents the jobs that are deployed in the past but still running at the beginning of the time step $t$, $B^{\text{cur}}(t)$ represents the jobs currently being considered that include time $t$ between their earliest start time and latest start time, and $B^{\text{fut}}(t)$ represents those jobs that have been submitted to the system by $t$ but have not reached their earliest start time so that will be considered in the future.


Thus the state of the system can be represented as
\begin{equation}
\label{eqn.state}
    s_{t} \leftarrow \left(B^{\text{his}}(t), B^{\text{cur}}(t),  B^{\text{fut}}(t), C_t\right)
\end{equation}

\subsubsection{Action}
The direct action to take at each time step $t$ for the problem is the chosen jobs among $B(t)$ to be deployed, which has a variable length because the set $B(t)$ is different at different time steps. Therefore, the direct action is
\begin{equation}
\label{eqn.ori_action}
a_t=\left(a^1_t,a^2_t,\ldots,a_t^{K_t}\right)\in\{0, 1\}^{K_t},  K_t = \left|B^{\text{cur}}(t)\right|
\end{equation}
where the length of this action vector at time $t$ is the cardinality of the set $B^{\text{cur}}(t)$, and for each item $a_t^j$ in this set $B^{\text{cur}}(t)$, $a_t^j=1$ means the job will be scheduled at time $t$ and $a_t^j = 0$ vice versa. However, to avoid the exploding action space for this discrete action setting, we use a continuous ``confidence score'' as the action and output the scheduling plan with these scores for all the jobs:
\begin{equation}
\mathit{CS}_t = \left(cs^1_t,cs^2_t,\ldots,cs^{K_t}_t\right)\in \mathbb{R}^{K_t},  K_t = \left|B^{\text{cur}}(t)\right|
\end{equation}
The jobs are then selected according to their confidence scores to transfer from $\mathit{CS}_t$ to $a_t$: sorting these job items according to their scores and scheduling them until reaching the capacity limit: for the sorted list $\mathit{CS}_t' = (cs^{(1)}_t,\ldots,cs^{(K_t)}_t)$, $cs^{(i)} \le cs^{(j)}$ for $1\le i<j\le K_t $, let $k$ to be the maximum number of jobs from the beginning of the list $(cs_t^{(1)},\ldots,cs_t^{(k)})$ with the sum of their required cores not exceeding the capacity limit, then let $a_t^{(i)} = 1$ for $1 \le i \le k$ and $a_t^{(i)} = 0$ for $k < i \le K_t$.

\subsubsection{Policy}
The stochastic policy for scoring the job candidates given the system state is fitted with a diagonal Gaussian distribution \cite{schulman2017proximal}:

\begin{equation}
\label{eqn.policy}
\mathit{CS}_t  \sim  \pi(\mathit{CS}_t|s_t) = N(\bm\mu_t, \bm\Sigma_t)
\end{equation}
where $\bm\mu_t = \left(\mu^{1}_t, \ldots, \mu^{K_t}_t \right)$ is the learned mean vector and $\bm\Sigma_t = \mathit{diag}\left(\sigma_t^1,\ldots,\sigma_t^{K_t}\right)$ is the learned diagonal standard deviation of the confidence score for the corresponding items obtained from the policy network respectively. 

\subsubsection{Reward}
The reward of the proposed method is set to be consistent with the system goal: to utilize the available capacity through the time efficiently and let the deferrable jobs run soon after they are ready. In order to combine the above two goals, the reward function is designed as the following form:
\begin{equation}
\label{eqn.reward}
    R^t = \sum_{i=1}^{K_t} \left(a_t^i  \cdot (r_i - \omega_1 p_i)\right) - \omega_2  v_t
\end{equation}
where $\omega_1$ and $\omega_2$ are the coefficients for the time delay penalty and the capacity violation penalty. 

\subsection{Network Architecture with Auxiliary Tasks}

 
 \subsubsection{Overall Structure}
To learn the policy in (\ref{eqn.policy}), we design a neural network for this specific model form with the structure displayed in Figure~\ref{fig:policy_net}. The job information and the available capacity in the system are formatted as (\ref{eqn.state}) before feeding into the network.  Besides, we extract the information from auxiliary tasks and concatenate it with each state vector in (\ref{eqn.state}).  Then, an encoding operation is conducted after this concatenation, followed by several Transformer encoder units. The output of the encoder module goes into two heads: policy head and value head. The output of the policy head is used to generate samples of the confidence scores for the jobs, and the value head outputs the corresponding value for the current state. The details of these components will be explained in the following content.


\begin{figure}[t]
    \centering
    \includegraphics[width=0.9\textwidth]{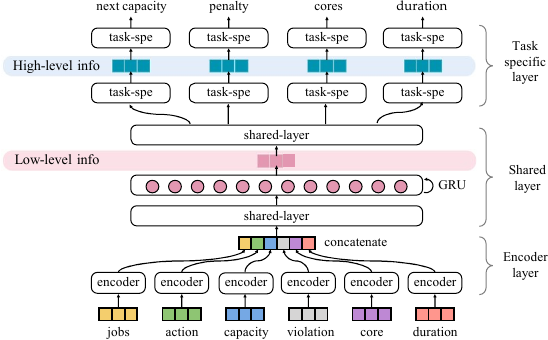}
    \caption{Illustration of the auxiliary tasks module.}
    \label{fig:mt_net}
    \vspace{-0.3cm}
\end{figure}

\subsubsection{Integration of Auxiliary Tasks}
To explicitly integrate the prediction of necessary knowledge of the system into the reinforcement learning model, we have considered four auxiliary tasks and add the information from these tasks into the state of the model: predicting for the next time step 1) the available capacity, 2)the average number of cores for the available jobs, 3) the average duration of the available jobs, and 4) the violation penalty caused by violating the available capacity.   The intuition behind this auxiliary tasks module is: we humans are able to make better scheduling plans with a better prediction of such auxiliary information. Therefore, by explicitly integrating the high-level and low-level information from the auxiliary prediction model, the state of the system can be better represented \cite{zintgraf2021deep}. 

The network structure of the auxiliary tasks module is shown in Figure \ref{fig:mt_net}. Past jobs, actions, capacity, violation status and some summary statistics including the average number of cores and average duration for the jobs are used to conduct the predictions of the auxiliary tasks. Here we use a Gate Recurrent Unit (GRU) in the prediction model, where other recurrent neural networks are also applicable. The shared layer after the GRU units would be separated and form a task specific layer. We extract both this shared layer after the GRU units and the task-specific layer as low-level and high-level information from these auxiliary tasks respectively, and these two layers are concatenated with the state to enrich it. Mean-square error loss is used to train these auxiliary tasks.

\subsubsection{Shared Encoder for Policy and Value Network}
After the processing of the state, an encoder module is used, which is shared by the policy network and the value network. This shared encoder module can enforce the sharing of information between the policy and the state value and help accelerate the training process. Inspired by \cite{vaswani2017attention}, our encoder layer consists of two sub-layers. The first one is a self-attention layer, and the second one is a position-wise fully connected feed-forward network. In each sub-layer, residual connection \cite{2016Deep} is employed followed by normalization layer. Thus, we can get a proper representation of each job from the encoder layer, which incorporates the self-information of the job and the correlation information with other jobs. 
For the policy head, we employ two position-wise fully connected feed-forward network whose output length is equal to the maximum number of input jobs of the network, followed by a layer of activation function, using $\tanh$ function and $\text{softplus}$ function as the activation function for mean $\bm\mu$ and variance $\bm\Sigma$ of the confidence score for each job respectively. 
For the value head, the output is calculated by averaging the output of a fully connected layer after the output of the encoder module, resulting in a one-dimensional state value $V(s)$ for each state.

\subsection{Model Training}
To train the proposed reinforcement learning model, we have adopted the proximal policy optimization (PPO) method proposed in \cite{schulman2017proximal} for better training efficiency and quality. 
It uses a truncated version of generalized advantage estimation, which is calculated as 
\begin{equation}
    A_t = \delta_t + (\gamma\lambda)\delta_{t-1} +... +  (\gamma\lambda)^{T-t}\delta_{T-1}
    \end{equation}
    where $\delta_t = r_t + \gamma V(s_{t+1}) - V(s_t)$ for the  $t^{\text{th}}$ step in a  trajectory segment with length $T$, $\gamma$ is the discount factor, and $\lambda$ is the exponential weight for generalized advantage estimation introduced in \cite{john2016gae}.
The surrogate loss function used in the policy network is 
\begin{equation}
\label{eqn.ppo_policy}
    L_t(\theta) = \min(R_t(\theta)A_t, \text{clip}(R_t(\theta), 1-\epsilon, 1+\epsilon)A_t)
\end{equation}
where the clip function clips $R_t(\theta)$ to be inside the interval $[1-\epsilon, 1+\epsilon]$ and $\epsilon$ is a hyperparameter with a small value.
Similarly, the surrogate loss for the value network is:
\begin{equation}
\label{eqn.ppo_value}
    L_t^{V} = (V_\theta(s_t) - V_t^{\text{targ}})^2
\end{equation}
where $V_t^{\text{targ}}$ is usually the calculated reward obtained from the last update iteration in implementation. The overall training process is shown in Algorithm \ref{alg:training}. 

\begin{algorithm}[tb]
\caption{Training Process of \OSDEC{}}
\label{alg:training}
\KwIn{Historical capacity $(C_1,...,C_T)$,\newline
    Job Requests $B(T)$, \newline
    Penalty parameters $\omega_1, \omega_2$, 
    Epoch number $E$,
    Max update iteration steps $L$\;
} 
\KwOut{The learned network parameters $\theta$\;}
\For{ $epoch \leftarrow 1$ \KwTo $E$}{
    \For {$traj \leftarrow 1$ \KwTo $L$}{
        \For {$t \leftarrow 1$ \KwTo $T$}{
            Pre-process the state $s_t$ as (\ref{eqn.state})\;
            Concatenate the vectors from auxiliary tasks with $s_t$\;
            Calculate $\bm\mu_t$, $\bm\Sigma_t$ and state value $V_t(s_t)$ through the network in Figure~\ref{fig:policy_net}\;
            Sample $CS_t$ according to ($\ref{eqn.policy}$)\;
            Obtain $a(t)$ according to $\mathit{CS}_t$\;
            Calculate $R^t$ according to (\ref{eqn.reward})\;
        }
    }
    Update the policy network by back-propogation with (\ref{eqn.ppo_policy})\;
    Update the value network by back-propogation with (\ref{eqn.ppo_value})\;
}
\Return{$\theta$}
\end{algorithm}


\section{Experiments}


\subsection{Dataset}
In order to validate the effectiveness of \OSDEC{}, we conduct extensive experiments on a latest public dataset that is collected from Microsoft Azure. The dataset is introduced in \cite{ori2020azuredata} and includes part of the workload on Azure Compute for 14 consecutive days. To make the dataset applicable for our algorithm and evaluation,  we have done the necessary data processing on it and will release the processed dataset.

\subsection{Competitors}
We consider the following two categories of methods as competitors. The first category is the classical heuristic methods, including  first-in-first-out (FIFO), shorted-job-first (SJF),  and Tetris \cite{grandl2014multi}. 
For the other category, 
we list six variants of our proposed method, including: 1) the variant using REINFORCE as the policy gradient algorithm instead of PPO; 2) the variant with fixed-length input (PPO), which uses a multi-layer perceptron neural network as the policy and value network; 3) the variant replacing the integrated neural network with the pointer network (PointerNet) to represent the state-of-the-art method for solving combinatorial optimization with RL; 4) PPO + Att, without the integration of the auxiliary tasks; 5) PPO + Att + $\textrm{Aux}_H$, using only high-level information from auxiliary tasks; 6) PPO + Att + $\textrm{Aux}_H$, using only low-level information from auxiliary tasks.

\subsection{Performance against competitors}
\subsubsection{Configuration}


All the competitors and our proposed method are implemented in the parallel learning architecture proposed in the previous section, involving 16 parallel workers.  The Adam optimizer with a linear decay learning rate is used to update all the parameters. We do the evaluation at each iteration, and each evaluation calculates the average reward over 100 trajectories without exploration noise. 


\subsubsection{Training Performance}

\begin{table}[t]
\centering
\setlength{\abovecaptionskip}{7pt}
\setlength{\tabcolsep}{15pt}
\caption{Scheduling performance of \OSDEC{} with competitors.}
\label{tab.perf}
\begin{threeparttable}
    \begin{tabular}{@{\hspace{6pt}}lccc@{\hspace{6pt}}}
    
    \toprule
    Algorithm & Utilization & Time Delay & Total Reward \\
    \midrule
    FIFO        & 3602.45           & -193.64           & 3408.82           \\
    SJF         & 3386.66           & -112.06           & 3274.66           \\
    Tetris      & 3689.65           & -282.42           & 3407.20           \\
    \midrule
    REINFORCE   & 3560.23 $\pm$ 62.88  & -121.64 $\pm$ 48.90  & 3301.06 $\pm$ 55.85  \\
    PPO         & 3596.25 $\pm$ 41.04  & -218.74 $\pm$ 37.46  & 3377.48 $\pm$ 39.21  \\
    PointerNet  & 3583.60 $\pm$ 49.48  & -205.76 $\pm$ 39.80  & 3377.91 $\pm$ 44.61  \\
    PPO+Att     & 3495.20 $\pm$ 45.52  & -95.66 $\pm$ 35.88   & 3387.22 $\pm$ 40.68  \\
    PPO+Att+Aux$_H$ & 3706.87 $\pm$ 36.34 & -101.28 $\pm$ 22.72 & 3444.00 $\pm$ 29.46 \\
    PPO+Att+Aux$_L$ & 3737.47 $\pm$ 39.76 & -240.42 $\pm$ 20.70 & 3466.22 $\pm$ 30.18 \\
    OSDEC       & \textbf{3761.24 $\pm$ 19.52} & \textbf{-250.25 $\pm$ 6.44} & \textbf{3475.37 $\pm$ 12.95} \\
    \bottomrule
    \end{tabular}
\end{threeparttable}
\end{table}

\begin{table}[t]
\centering
\setlength{\abovecaptionskip}{7pt}
\setlength{\tabcolsep}{18pt}
\caption{Scheduling performance under different settings.}
\begin{threeparttable}
    \begin{tabular}{@{\hspace{10pt}}lcc@{\hspace{10pt}}}
    \toprule
    Algorithm\hspace{-3pt}              & Deferrable   & Real-time    \\ 
    \midrule
                FIFO                    & 3611.01        & 3449.22 \\
                 SJF                    & 3398.76         & 3298.38         \\
                
                 Tetris                 & 3696.79         & 3503.01      \\
    
    \midrule
                REINFORCE               & 3567.80$\pm$69.74        & 3360.28$\pm$48.82 \\
                PPO                     & 3589.87$\pm$45.10        & 3384.50$\pm$36.08  \\
                PointerNet              & 3576.02$\pm$57.41        & 3398.82$\pm$45.82  \\
                PPO+Att                 & 3525.43$\pm$45.55        & 3340.77$\pm$16.53 \\
                PPO+Att+Aux$_H$         & 3708.02$\pm$25.22        & 3411.08$\pm$20.87   \\
                PPO+Att+Aux$_L$         & 3744.02$\pm$25.05        & 3442.96$\pm$19.88  \\
                OSDEC                   & \textbf{3781.17$\pm$22.25}        & \textbf{3521.23$\pm$15.31} \\
               
    \bottomrule
    \end{tabular}
\end{threeparttable}
\label{tab.deferrableFactor}
\end{table}

\begin{wrapfigure}{r}{0.45\textwidth}
  \centering
  \includegraphics[width=0.45\textwidth]{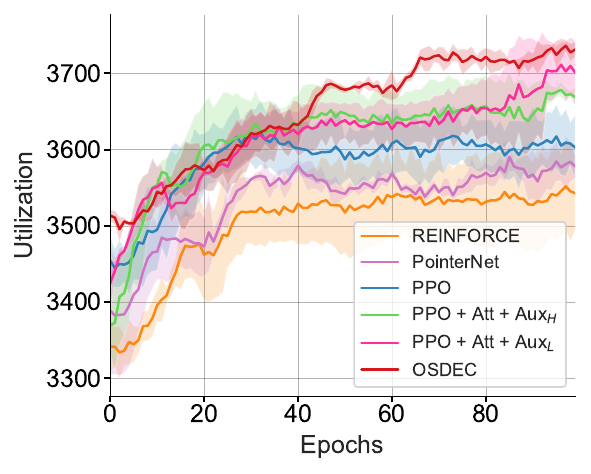}
  \caption{Training performance curve.}
  \label{fig:training_performance}
\end{wrapfigure}

The training performance of our proposed method against other variants is shown in Figure \ref{fig:training_performance}. 
It is shown that the proposed method achieves the best performance and has the least variance among different runs. Without either high-level or low-level auxiliary information, the performance would decrease.

The REINFORCE method performs worse than the other methods, because the REINFORCE policy gradient method can lead to a sub-optimal solution and suffers from large variance as reported in the related literature \cite{zhao2011analysis}.  Although PPO has better performance than other two methods, the proposed \OSDEC{} method can outperform PPO by adding attention mechanism and information from auxiliary tasks. Moreover, the PointerNet method also  shows a poor performance because its encoding structure is more suitable for temporal sequence rather than the unordered job set in this scenario.

\setlength{\intextsep}{-1pt}
\captionsetup[wrapfigure]{skip=-5pt}


\subsubsection{Scheduling Performance}

We also compare the scheduling performance of our proposed method with other competitors on the public Azure dataset. Specifically, the total reward, the utilization and the job deployment time delay penalty are reported.  Overall 100 trajectories are sampled to perform the heuristic methods and train the RL models, and we calculate the three metrics over the 100 trajectories as well as their standard deviations.  Table \ref{tab.perf} shows the detailed results. Note that for heuristic methods there is no standard deviation because they are deterministic.

From the results in Table \ref{tab.perf}, the proposed \OSDEC{} method shows better performance against all the other methods in total reward.  Moreover, the utilization of \OSDEC{} is the highest among all the compared algorithms, which verifies the effectiveness of the proposed method.  Our designed modules, including the attention mechanism, the high-level information and low-level information from auxiliary tasks, all show their own improvements on the performance of the scheduling task. The Tetris heuristic can perform quite well in the utilization than other heuristic methods, but also has the largest time delay penalty.

\subsubsection{Benefits of Deferrable Setting}
To investigate whether our proposed method actually benefits from the deferrable setting which allows it to foresee future jobs for better planning, we compare the performance of our proposed method as well as other methods under the deferrable scheduling setting with these methods in the real-time scheduling setting. In the real-time setting, there's no valid window for all the submitted jobs, and all the jobs should be deployed instantaneously upon submission. In both these settings, the goal has been set to maximize the utilization for all the methods. The results of this comparison are shown in Table \ref{tab.deferrableFactor}. By allowing the jobs to be deferrable, all the methods can achieve a higher utilization compared to the real-time deployment setting. Moreover, our proposed method achieves significant improvement under the deferrable setting compared to the real-time setting, which shows the benefits of the deferrable setting.




\section{Conclusion}
In this paper, we propose \OSDEC{}, a deep reinforcement learning based method for online scheduling of deferrable jobs in cloud computing. In the proposed model, the information of pre-collected jobs is well incorporated into the state to help the model make better scheduling plans under variable available capacity. To explicitly incorporate the prediction related information into our method, we design several auxiliary prediction tasks to enrich the state.  Moreover, we design a network architecture with Transformer encoder shared by both policy network and value network to effectively and efficiently learn the scheduling policy. Extensive experiments have been conducted to validate the effectiveness of the proposed method on a public dataset from a real-world cloud computing platform. 
For future work, we plan to extend this reinforcement learning framework with auxiliary tasks to other scheduling problems with different settings. Furthermore, it is worthwhile to consider how to apply safe reinforcement learning methods for more robust scheduling performance when deploying in cloud computing platforms.

\bibliographystyle{plain}
\bibliography{neurips_2024}

\begin{thebibliography}{10}

\bibitem{dongpredictive21}
Hang Dong, Boshi Wang, Bo~Qiao, Wenqian Xing, Chuan Luo, Si~Qin, Qingwei Lin,
  Dongmei Zhang, Gurpreet Virdi, and Thomas Moscibroda.
\newblock Predictive job scheduling under uncertain constraints in cloud
  computing.
\newblock In {\em IJCAI}, pages 3627--3634, 2021.

\bibitem{Fanjul2017}
Luis Fanjul-Peyro, Federico Perea, and Rubén Ruiz.
\newblock Models and matheuristics for the unrelated parallel machine
  scheduling problem with additional resources.
\newblock {\em European Journal of Operational Research}, 260(2):482 -- 493,
  2017.

\bibitem{Fleszar2018}
Krzysztof Fleszar and Khalil~S. Hindi.
\newblock Algorithms for the unrelated parallel machine scheduling problem with
  a resource constraint.
\newblock {\em European Journal of Operational Research}, 271(3):839 -- 848,
  2018.

\bibitem{ghodsi2011dominant}
Ali Ghodsi, Matei Zaharia, Benjamin Hindman, Andy Konwinski, Scott Shenker, and
  Ion Stoica.
\newblock Dominant resource fairness: Fair allocation of multiple resource
  types.
\newblock In {\em NSDI}. {USENIX} Association, 2011.

\bibitem{grandl2014multi}
Robert Grandl, Ganesh Ananthanarayanan, Srikanth Kandula, Sriram Rao, and
  Aditya Akella.
\newblock Multi-resource packing for cluster schedulers.
\newblock {\em Computer Communication Review}, 44(4):455--466, 2014.

\bibitem{guo2020cloud}
Wenxia Guo, Wenhong Tian, Yufei Ye, Lingxiao Xu, and Kui Wu.
\newblock Cloud resource scheduling with deep reinforcement learning and
  imitation learning.
\newblock {\em IEEE Internet of Things Journal}, 8(5):3576--3586, 2020.

\bibitem{ori2020azuredata}
Ori Hadary, Luke Marshall, Ishai Menache, Abhisek Pan, Esaias~E Greeff, David
  Dion, Star Dorminey, Shailesh Joshi, Yang Chen, Mark Russinovich, and Thomas
  Moscibroda.
\newblock Protean: {VM} allocation service at scale.
\newblock In {\em OSDI}, pages 845--861, 2020.

\bibitem{2016Deep}
Kaiming He, Xiangyu Zhang, Shaoqing Ren, and Jian Sun.
\newblock Deep residual learning for image recognition.
\newblock In {\em CVPR}, pages 770--778. IEEE, 2016.

\bibitem{isard2009quincy}
Michael Isard, Vijayan Prabhakaran, Jon Currey, Udi Wieder, Kunal Talwar, and
  Andrew Goldberg.
\newblock Quincy: fair scheduling for distributed computing clusters.
\newblock In {\em SIGOPS}, pages 261--276. ACM, 2009.

\bibitem{kartal2019terminal}
Bilal Kartal, Pablo Hernandez-Leal, and Matthew~E Taylor.
\newblock Terminal prediction as an auxiliary task for deep reinforcement
  learning.
\newblock In {\em AIIDE}, number~1, pages 38--44, 2019.

\bibitem{kumar2021implicit}
Aviral Kumar, Rishabh Agarwal, Dibya Ghosh, and Sergey Levine.
\newblock Implicit under-parameterization inhibits data-efficient deep
  reinforcement learning.
\newblock In {\em ICLR}, 2021.

\bibitem{latip2011highest}
Rohaya Latip and Zulkhairi Idris.
\newblock Highest response ratio next (hrrn) vs first come first served (fcfs)
  scheduling algorithm in grid environment.
\newblock In {\em ICSECS}, pages 688--693. Springer, 2011.

\bibitem{li2016negotiation}
Ji~Li, Yanzhi Wang, Xue Lin, Shahin Nazarian, and Massoud Pedram.
\newblock Negotiation-based resource provisioning and task scheduling algorithm
  for cloud systems.
\newblock In {\em ISQED}, pages 338--343. {IEEE}, 2016.

\bibitem{Lin2019NIPS}
Xingyu Lin, Harjatin Baweja, George Kantor, and David Held.
\newblock Adaptive auxiliary task weighting for reinforcement learning.
\newblock In {\em NeurIPS}, 2019.

\bibitem{liu2017hierarchical}
Ning Liu, Zhe Li, Jielong Xu, Zhiyuan Xu, Sheng Lin, Qinru Qiu, Jian Tang, and
  Yanzhi Wang.
\newblock A hierarchical framework of cloud resource allocation and power
  management using deep reinforcement learning.
\newblock In {\em ICDCS}, pages 372--382. IEEE, 2017.

\bibitem{lyle_effect2021}
Clare Lyle, Mark Rowland, Georg Ostrovski, and Will Dabney.
\newblock On the effect of auxiliary tasks on representation dynamics.
\newblock In {\em AISTATS}. PMLR, 2021.

\bibitem{mao2016resource}
Hongzi Mao, Mohammad Alizadeh, Ishai Menache, and Srikanth Kandula.
\newblock Resource management with deep reinforcement learning.
\newblock In {\em HotNets '16}, pages 50--56. ACM, 2016.

\bibitem{mesbahi2018reliability}
Mohammad~Reza Mesbahi, Amir~Masoud Rahmani, and Mehdi Hosseinzadeh.
\newblock Reliability and high availability in cloud computing environments: a
  reference roadmap.
\newblock {\em Human-centric Computing and Information Sciences}, 8(1):1--31,
  2018.

\bibitem{Narwal2018FIFO}
Abhikriti Narwal and Sunita Dhingra.
\newblock Enhanced task scheduling algorithm using multi-objective function for
  cloud computing framework.
\newblock In {\em NGCT}, pages 110--121. Springer Singapore, 2018.

\bibitem{john2016gae}
John Schulman, Philipp Moritz, Sergey Levine, Michael~I. Jordan, and Pieter
  Abbeel.
\newblock High-dimensional continuous control using generalized advantage
  estimation.
\newblock In {\em ICLR}, 2016.

\bibitem{schulman2017proximal}
John Schulman, Filip Wolski, Prafulla Dhariwal, Alec Radford, and Oleg Klimov.
\newblock Proximal policy optimization algorithms.
\newblock {\em CoRR}, abs/1707.06347, 2017.

\bibitem{strens2006combining}
Malcolm~JA Strens.
\newblock Combining stochastic task models with reinforcement learning for
  dynamic scheduling.
\newblock In {\em ICAPS}, pages 426--429, 2006.

\bibitem{vaswani2017attention}
Ashish Vaswani, Noam Shazeer, Niki Parmar, Jakob Uszkoreit, Llion Jones,
  Aidan~N. Gomez, Lukasz Kaiser, and Illia Polosukhin.
\newblock Attention is all you need.
\newblock In {\em NeurIPS}, pages 5998--6008, 2017.

\bibitem{williams1992simple}
Ronald~J Williams.
\newblock Simple statistical gradient-following algorithms for connectionist
  reinforcement learning.
\newblock {\em Machine learning}, 8(3):229--256, 1992.

\bibitem{zhao2011analysis}
Tingting Zhao, Hirotaka Hachiya, Gang Niu, and Masashi Sugiyama.
\newblock Analysis and improvement of policy gradient estimation.
\newblock In {\em NeurIPS}, 2011.

\bibitem{zhou2021deep}
Guangyao Zhou, Wenhong Tian, and Rajkumar Buyya.
\newblock Deep reinforcement learning-based methods for resource scheduling in
  cloud computing: {A} review and future directions.
\newblock {\em CoRR}, abs/2105.04086, 2021.

\bibitem{zintgraf2021deep}
Luisa Zintgraf, Sam Devlin, Kamil Ciosek, Shimon Whiteson, and Katja Hofmann.
\newblock Deep interactive bayesian reinforcement learning via meta-learning.
\newblock In {\em AAMAS}, pages 1712--1714, 2021.

\end{thebibliography}

\appendix
\onecolumn

\section{Detailed Experiment Settings}
\begin{table}[h!] 
\centering
\setlength{\abovecaptionskip}{7pt}
\setlength{\tabcolsep}{11pt}
\caption{Detailed experiment setting. }
\begin{threeparttable}[h]
\begin{tabular}{l@{}c}
\toprule
Hyperparameter & Value \\
\midrule
Batch size & 2048\\
Mini-batch size & 64\\
Policy learning rate & 1e-4 $\rightarrow$ 1e-5 \\
Value learning rate & 2e-4 $\rightarrow$ 2e-5 \\
Discount factor $\gamma$ & 0.99\\
Time delay penalty coefficient $\omega_1$ & 2\\
Capacity violation penalty coefficient $\omega_2$ & 10\\
\#Attention head & 1\\
Policy mean activation function & $\tanh{}$\\
Policy variance activation function & $\text{softplus}$\\
Multi-Task learning rate & 1e-2 $\rightarrow$ 1e-3 \\
High prior information coefficient & 10 \\
Low prior information coefficient & 10 \\
High prior information dimension & 5 \\
Low prior information dimension & 5 \\
\bottomrule
\end{tabular}
\end{threeparttable}
\label{table.hyper}
\end{table}

\vspace{10pt}
\section{Sensitivity Study of Environment Settings}

\begin{table}[h!]
\centering

\setlength{\abovecaptionskip}{7pt}
\caption{Sensitivity study of environment settings.}
\begin{threeparttable}[h]
\begin{tabular}{lrrrrrrr}
\toprule
\multirow{2}{*}{Algorithm} &  \multicolumn{3}{c}{Core distribution\tnote{*}}  & \multicolumn{3}{c}{Duration Distribution\tnote{\#}} \\
\cmidrule(rr){2-4} \cmidrule(rr){5-7}
                &  Core case 1  & Core case 2 & Core case 3 & Dur Case 1 & Dur Case 2 & Dur Case 3\\
\midrule

            FIFO                & 3651.49& 7469.07 & 5009.48 & 8693.60 & 5536.08 & 3467.29 \\
            SJF                 & 3518.63& 5771.45 & 4412.64 & 7633.65 & 5032.28 & 3340.02 \\
            Tetris              & 3856.63& 9144.62 & 5537.23 & 9475.69 & 5923.69 & 3633.64 \\
\midrule
            REINFORCE           & 3613.78& 7148.49 & 4869.69 & 8136.24 & 5327.90 & 3399.00 \\
            PPO                 &3648.38 & 7415.11 & 4972.13 & 8324.38 & 5386.61 & 3427.17 \\
            PointerNet          &3649.47 & 7126.47 & 4859.10 &  8281.23&  5388.20& 3442.37 \\
            PPO + Att           &3662.41 & 7957.08 & 4663.42 & 8248.63 & 5472.42 & 3357.90 \\
            PPO + Att + Aux$_H$ & 3707.01& 8425.30 & 5096.22 & 8571.09 & 5592.46 & 3536.01 \\
            PPO + Att + Aux$_L$ &3781.08 & 8560.46 & 5430.28 & 9113.23 &  5876.61& 3595.21 \\
            OSDEC               &3884.28 & 9200.82 & 5565.89 & 9593.68 & 5853.81 & 3631.21 \\

\bottomrule
\end{tabular}
\begin{tablenotes}
     \vspace{5pt}
     \item[*] Core case 1: $[0.51, 0.37, 0.08, 0.04]$, Core case 2: $[0.25, 0.25, 0.25, 0.25]$, Core case 3: $[0.4, 0.3, 0.2, 0.1]$.
     \item[\#] Dur Case 1: $[1/6, 1/6, 1/6, 1/6, 1/6, 1/6]$, Dur Case 2: $[0.4, 0.25, 0.15, 0.1, 0.05, 0.05]$, Dur Case 3: $[0.6, 0.25, 0.1, 0.05, 0, 0]$.
\end{tablenotes}
\end{threeparttable}
\label{tab.ablation_env}
\end{table}

\end{document}